\documentclass[twocolumn,floatfix,amsmath,amssymb,superscriptaddress]{revtex4-1}
\usepackage{graphicx}

\usepackage{hyperref}
\hypersetup{
bookmarks=false,
pdftitle={q-K-theory}
anchorcolor=blue,
colorlinks=true,
citecolor=blue,
urlcolor=blue,
linkcolor=blue}
\usepackage{marginnote}
\usepackage[author=Juanjo]{pdfcomment}
\usepackage{xcolor}

\usepackage{bm}
\usepackage{amsfonts}

\begin{document}

\title{Quantum Navigation and Ranking in Complex Networks}

\author{Eduardo S\'anchez-Burillo}
\affiliation{Instituto de Ciencia de Materiales de Arag\'on (ICMA), CSIC-Universidad de Zaragoza, E-50012 Zaragoza, Spain.}
\affiliation{Departamento de F\'{\i}sica de la Materia Condensada,
  University of Zaragoza, Zaragoza 50009, Spain}

\author{Jordi Duch}
\affiliation{Departament d'Enginyeria Inform{\`a}tica i
  Matem{\`a}tiques, Universitat Rovira i Virgili, 43007 Tarragona,
  Spain}

\author{Jes{\'u}s G{\'o}mez-Garde\~{n}es}
\affiliation{Departamento de F\'{\i}sica de la Materia Condensada,
  University of Zaragoza, Zaragoza 50009, Spain}
\affiliation{Institute for Biocomputation and Physics of Complex
Systems (BIFI), University of Zaragoza, Zaragoza 50018, Spain}

\author{David Zueco}
\affiliation{Instituto de Ciencia de Materiales de Arag\'on (ICMA), CSIC-Universidad de Zaragoza, E-50012 Zaragoza, Spain.}
\affiliation{Departamento de F\'{\i}sica de la Materia Condensada,
  University of Zaragoza, Zaragoza 50009, Spain}
\affiliation{Fundaci\' on ARAID, Paseo Mar\'{\i}a Agust\'{\i}n 36, E-50004 Zaragoza, Spain.}

\date{\today}

\begin{abstract}
Complex networks are formal frameworks capturing the interdependencies between the elements of large systems and databases. This formalism allows to use network navigation methods to rank the importance that each constituent has on the global organization of the system. A key example is Pagerank navigation which is at the core of the most used search engine of the World Wide Web. Inspired in this classical algorithm, we define a quantum navigation method providing a unique ranking of the elements of a network. We analyze the convergence of quantum navigation to the stationary rank of networks and show that quantumness decreases the number of navigation steps before convergence. In addition, we show that quantum navigation allows to solve degeneracies found in classical ranks. By implementing the quantum algorithm in real networks, we confirm these improvements and show that quantum coherence unveils new hierarchical features about the global organization of complex systems.
\end{abstract}

\maketitle

\section{Introduction}

The search for information in the World Wide Web (WWW) through search engines has turned into a daily habit and an essential tool to fulfill most of our work duties. An ideal search engine looks for the information the user is querying amongst billions of webpages in real time, and produces a ranking of the results sorted according the user expectations. Although not being among the first search engines available, the Google search engine was the first to achieve these goals efficiently, establishing one of the milestones of the digital era. Its main novelty was to classify and rank webpages based on the interrelations created between them through the hyperlinks\cite{Marchiori}, rather than using only their intrinsic features (such as the page content). Google's ranking algorithm, known as Pagerank\cite{PageBrin} (PR), is rooted in a diffusion process that mimics the user's navigation through webpages as the motion of a random walker following hyperlink pathways. 

After Google's global success, multiple applications of PR navigation have blossomed whenever a large dataset can be mapped into a complex network encoding the interdependencies between items. Some examples include the classification of species within ecosystems\cite{Allesina} or the evaluation of scientists' impact in different research disciplines\cite{Santo}. In these examples PR has successfully produced a classification in which the importance of each element accounts for its status within the whole system. Recent studies\cite{BarabasiNC} have shown that what underpins the success of PR navigation in the classification of network elements according to their global role is the scale-free nature of most real networks\cite{rev:albert,rev:newman,rev:bocc}, highlighting the influence that the structural {properties} of real networks have on the collective outcome of the dynamical processes taking place on top of them\cite{vespinature,watts,szabo,RMPsoc,reviewanxo,PRsync,rev:doro2}.

Very recently, the dynamical setting of network-based diffusion has been extended to the quantum domain\cite{Muelken2011}. In this framework, quantum random walks have already shown their potential for practical purposes as they offer a quadratic speed up with respect to the classical algorithms for searching in unsorted databasis\cite{Kempe2003, Kendon2006}. Considering these recent advances, it is tempting to explore the potential application of quantum random walks for ranking elements in large complex networks, {\it i.e.}, a quantum rank (QR) algorithm, that improves PR based on the benefits that quatumness may introduce\cite{NielsenChuang}. To achieve this challenge we need to solve two important issues, {\it (i}) how to extend the quantum formulation of a walk in a similar way as PR did with classical walks so to find a reliable and unique ranking reflecting the topology of the network and {\it (ii)} how to identify the role that quantum coherence plays to capture new features about the organization of large complex networks.
% and to resolve non-local relations among their constituents. 

In the present work we show how to design a QR navigation as an hybrid classical-quantum walk. Our theoretical approach relies on a master equation for the walkers motion interpolating between purely quantum coherent dynamics and the classical diffusion\cite{Aspuru}. We explicitly show that a suitable choice of the master equation for the walkers motion allows to find a stationary solution yielding a unique and reliable QR.
We show that the interplay of quantum coherence with the classical hopping dynamics yields an optimal operation point so that QR navigation becomes remarkably faster than its classical counterpart. We will further analyze the effects that quantum coherence produces by comparing the rankings produced by QR and PR navigations. As a result we show that QR is able to split the degeneracies found in the ranking of PR and, more importantly, to unveil hidden hierarchies in the rank of lowly connected items that escape the resolution of PR.
 
\section*{Results}

\subsection{Quantum {\it versus} Classical Random Walks in complex networks.}

The architecture of a complex network is described by means of a collection of $N$ nodes and $L$ edges accounting for the elements of
the system and their pairwise interactions respectively. These latter interactions are encoded into a $N\times N$ adjacency matrix ${\cal A}=\{a_{ij}\}$, so that $a_{ij}=1$ when a link from $i$ to $j$ exists while $a_{ij}=0$ otherwise. Given this simple formulation one can design a variety of ways for exploring the network topology, being a random walk the most simple way to navigate networks. 

A random walk is usually described as a time-discrete process. At each time step, a walker hops from a node $i$ to another node $j$ provided such connection exists, {\it i.e.}, $a_{ij}=1$. The dynamics can be formulated in terms of a transition matrix ${\Pi}=\{\pi_{ij}\}$ being the entry $\pi_{ij}$ the probability that the walker jumps from $i$ to $j$. For usual random walks the transition matrix $\Pi$ is defined as $\Pi_{ij} = a_{ij}/k_i$,  where $k_i=\sum_{j=1}^N a_{ij}$ is the out-degree of the node $i$, {\it i.e.}, the number of links outgoing from node $i$.  The dynamics of the random walk is monitored by the time evolution of a vector  ${\bf p}^t=\{p_i^t\}$  whose $i$-th component accounts for the probability that the walker is placed at node $i$ at time $t$. Then, starting from some initial occupation probability, ${\bf p}^0$, the dynamics of the walker follows the time-discrete rate equation:
\begin{equation}
p_i^{t} = \sum_{j=1}^N \Pi_{ij} p_j^{t-1}\;,\;\; {\mbox{with}}\; i=1,...,N\;.
\label{eq:CRWdisc}
\end{equation}
The time-continuum version of a classical random walk can be also easily obtained from the above rate equation as: 
\begin{equation}
\label{crw}
\partial_t  \vec p = ( \Pi - \mathbb {I}) \vec p\;,
\end{equation}
where $\mathbb {I}$ is the identity matrix. On both cases, the iteration of Eq. (\ref{eq:CRWdisc}) and Eq. (\ref{crw}) ends up in a stationary state ${\bf p}^{*}=\lim_{t\rightarrow \infty} {\bf p}^{t}$. This stationary solution ${\bf p}^{*}$ is the cornerstone of any diffusion-based ranking as each of its components $p_i^*$  accounts for the importance of the corresponding node.

The usual map between classical and quantum random walks consists in identifying the jump probabilities (contained in $\Pi$ in the classical formulation) with ladder operators of a tight-binding Hamiltonian whereas walkers are described by two level systems, {\it i.e.}, qubits\cite{Muelken2011}.   Quantum walks have been discussed in both continuum and discrete versions ones. However, since they were originally conceived as implementations of quantum algorithms, all of them  share an important problem regarding the design of a QR: as quantum walks are reversible, they do not have a stationary solution for the occupation probabilities, thus making the definition of QR in these schemes quite challenging\cite{Paparo2011} (see Methods for a deeper discussion).

Here we take a different route. In analogy to the classical Markovian dynamics for the occupation probabilities defined in Eq. (\ref{crw}), we write its quantum analogue considering the  $N \times N$ quantum density matrix $\varrho$. In this case, the map between quantum and classical random walks lies on the fact that the diagonal elements of $\varrho$ account for the occupation probabilities of each node: $\langle i | \varrho | i \rangle = \varrho_{ii} = p_i$. Following this prescription, it turns out that any Markovian quantum evolution can be written in the form\cite{Rivas} (see Methods):
\begin{align}
\label{lindblad}
 \frac{d \varrho}{dt}= & -{\rm i}(1-\alpha)[H,\varrho]
\\ \nonumber
&+\alpha\sum_{(i,j)} \gamma_{(i,j)} \left(L_{(i,j)}\varrho L_{(i,j)}^\dagger-\frac{1}{2}
\{L_{(i,j)}^\dagger L_{(i,j)},\varrho \}\right)\;,
\end{align}
where $H$ is the Hamiltonian ($H = H^\dagger$) incorporating the (quantum) coherent dynamics,  whereas $\gamma_{(i,j)} >0$ and the operators $\{L_{(i,j)}\}$ (the so-called dissipators) are the ones responsible of irreversibility. The parameter $\alpha\in [0,1]$ quantifies the interplay between unitary and irreversible dynamics. The above formulation avoids the aforementioned problem associated to unitary evolution, {\it i.e.}, the absence of a stationary distribution for the quantum occupation probabilities. Moreover, the irreversible part of Eq. (\ref{lindblad}) can accommodate any classical random walk by choosing the definition of the dissipators. For instance, the usual random walk can be incorporated by choosing: $L_{(i,j)} =|i\rangle \langle j |$ and $\gamma_{(i,j)}= \Pi_{ij}$. In this case,  in the limit $\alpha =1$, the equations for the diagonal elements, $\varrho_{jj} = p_j$, meet those of the usual classical walk, Eq. (\ref{crw}).  Thus, Eq. (\ref{lindblad}) interpolates between classical ($\alpha=1$) and purely quantum random walks ($\alpha=0$).  Between both limits the dynamics incorporates the interplay between classical hopping and quantum coherence thus making navigation sensitive to non-local correlations between the nodes and paving the way to a more global ranking. 

\subsection{Quantum rank.} 

As mentioned previously, once the network navigation is defined the ranking of its nodes is done by evaluating each of the components of the stationary distribution, ${\bf p}^{*}$. This stationary distribution corresponds to the eigenvector of the  transition matrix $\Pi$ with (the largest) eigenvalue $1$,  $\Pi {\bf p}^{*} = {\bf p}^{*}$. However, the existence and unicity of ${\bf p}^{*}$ is only guaranteed provided the transition matrix fulfills the Perron-Frobenius theorem, {\it i.e.}, it is irreducible and aperiodic\cite{blanchard}. In order to satisfy this latter condition in any kind of network topology, the PR navigation incorporates a long-distance hopping probability to the usual classical walk. This ingredient  transforms the usual transition matrix, $\Pi$, into the so-called Google matrix:
\begin{equation}
\label{Gmatrix}
G = q \Pi + (1-q) F \;,
\end{equation}
where $q\in[0,1]$ is a parameter whose value is typically set to $q=0.9$ and $F$ is the long-distance hopping matrix, $F_{ij}=1/(N-1)$ if $i\neq j$ while $F_{ii}=0$ otherwise. By doing so, the transition matrix $G$ has a unique eigenvector with maximum eigenvalue $1$.  
Thus, the ranking of websites is related to the components of the eigenvector corresponding to the (unique) eigenvalue $1$ of the Google matrix, Eq. (\ref{Gmatrix}).  
  
In the quantum case we face a similar problem. The definition of the dissipators in Eq. (\ref{lindblad}) with the form of the classical operators, $L_{(i,j)} =|i\rangle \langle j |$  and $\gamma_{(i,j)}=\Pi_{ij}$, does not guarantee a unique stationary distribution for any network topology.
\color{black} 
 However, we prove in the Methods section  (see Theorem) that, when dissipators are defined as $L_{(i,j)}= |i\rangle \langle j |$ with $\gamma_{(i,j)}=G_{ij}$ (the Google matrix), 
\color{black}
Eq. (\ref{lindblad}) has always a unique stationary solution. In a nutshell, the patching mechanism used in the Google matrix is generalized here to the quantum case in a natural way. As a consequence, by incorporating the PR navigation in the irreversible part of Eq. (\ref{lindblad}) we can compute a reliable QR that incorporates the effects of coherences in the navigation.

\subsection{Quantum convergence.}  

\begin{figure}
\includegraphics[width=1.\columnwidth]{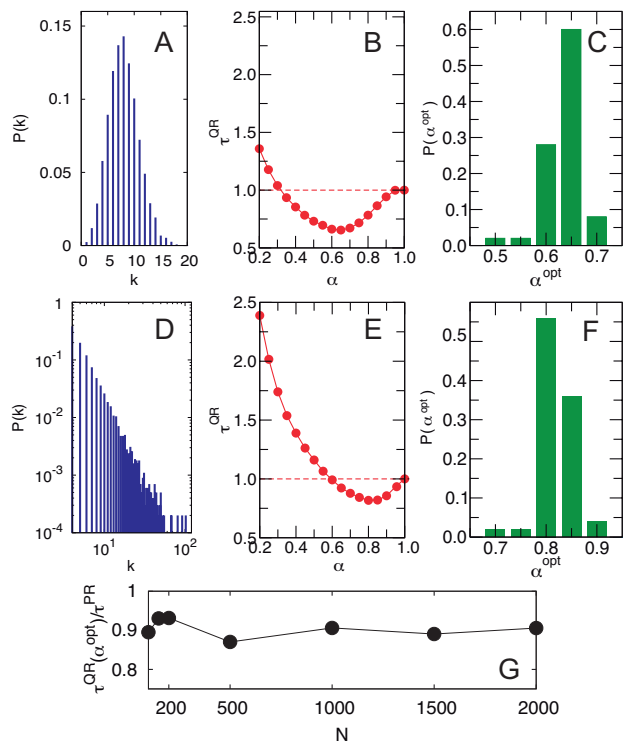}
\caption{{\bf Convergence times: QR {\it versus} PR.} In this figure we show the evolution of the convergence time to the stationary solution of QR, $\tau^{QR}$, as a function of the quantum-classical interplay, $\alpha$, introduced in the QR navigation. In the top part of the figure we focus on ER graphs in which the probability, $P(k)$, that a node is connected to $k$ other nodes (the degree distribution) follows a Poisson distribution as show in panel {\bf (a)}. Panel {\bf (b)} shows the averaged evolution $\tau^{QR}(\alpha)$ compared to the convergence time of PR, $\tau^{PR}$, in ER graphs of $N=200$ nodes whereas panel {\bf (c)} shows the probability that a value $\alpha^{opt}$ is obtained for a single network. It is clear that QR reaches faster the stationary solution (the final rank) for a broad range of $\alpha$ values, obtaining the maximum outperformance for $\alpha^{opt}\simeq 0.65$. The architecture of SF networks is rather different as shown in their power-law degree distribution,  see panel {\bf (d)}. In panels {\bf (e)} and {\bf (f)} we plot the evolution $\tau^{QR}(\alpha)/\tau^{PR}$ in SF networks with $P(k)\sim k^{-3}$ and $N=200$ nodes and the corresponding distribution of $\alpha^{opt}$ values obtained respectively. We observe again that for a range of $\alpha$ values QR converges faster than PR to a stationary ranking, reaching the maximum outperformance at $\alpha^{opt}\simeq 0.8$. Finally, in panel {\bf (g)} we show, for SF networks, the scaling of the ratio $\tau^{QR}/\tau^{PR}$ at $\alpha^{opt}$ with the size of the network, $N$. As observed, the former ratio remains roughly constant, thus showing the robustness of the results shown in the previous panels.}
\label{fig:tau}
\end{figure}

We now study the performance of our QR compared with that of PR. First we focus on one important aspect for the computation of a rank in large networks: the convergence time, $\tau$, to the stationary state of the navigation dynamics, 
${\varrho}^*$, {\it i.e.}, the rank. Being linear, the time evolution for the master equation (\ref{lindblad}) can be written in a compact form as $\dot \varrho = {\mathcal L} [\varrho]$.  Consequently, the evolution is determined by the spectrum of ${\mathcal L}$ and the convergence to the stationary state is upper-bounded by
$\tau =  |{\rm Re} [\lambda_1] \, |^{-1}$, with $\lambda_1$ being the first non-zero eigenvalue of ${\mathcal L}$.   
%As argued in the Methods section, we compute the actual convergence time by solving the equation (\ref{lindblad}), starting from the usual initial condition, $\varrho_{ij} = 1/N \delta_{ij}$.

To study the convergence time of QR we have considered two of the most paradigmatic network topologies: Erd\"os-R\'enyi (ER) and scale-free (SF) networks. These two networks differ in the distribution $P(k)$ of the number of contacts (degree), $k$, per node. In ER graphs most of the nodes share the same degree and $P(k)$ is shown to be a Poisson distribution (see Fig. 1.a). However, real-world networks follow a SF pattern characterized by a power-law degree distribution: $P(k)\sim k^{-\gamma}$ (see Fig. 1.b) where the exponent lies in the interval $2<\gamma<3$ so that the variability of the degrees found in the network becomes extremely large. 
%\sout{(since $\langle k^2\rangle$ diverges in the thermodynamics limit) due to the presence of largely connected nodes (the hubs).}
 
In Figs. \ref{fig:tau}1.b and 1.e we plot the evolution of  $\tau$ as a function of the classical-quantum interplay included in the QR navigation [quantified by $\alpha$ in Eq. (\ref{lindblad})] for SF and ER networks respectively. We observe that for $\alpha \to 0$ (approaching the fully-quantum limit) the value of $\tau$ diverges, as expected, since in this limit the dynamics becomes fully-unitary. On the other hand, in the classical limit, $\alpha \to 1$, the convergence time $\tau^{QR}$ tends to its classical counterpart, reaching $\tau^{PR}$ at $\alpha=1$.  The striking result appears for intermediate values of $\alpha$: the curve $\tau(\alpha)$ shows a minimum for some $0<\alpha^{opt}<1$. Thus, there exists an optimal value $\alpha^{opt}$ pointing out that an hybrid classical-quantum navigation outperforms significantly the time needed for ranking with PR. {\color{black} In Figs. \ref{fig:tau}1.c and 1.f we plot an histogram showing the probability of finding a given value of $\alpha^{opt}$, obtained after navigating an ensemble of $50$ SF and ER networks of $N=200$, respectively.} This enhancement resembles the {\it quantum stochastic resonance}-like phenomenon\cite{Grifoni96} previously found in other contexts, such as entanglement generation\cite{Huelga07} or efficient light-harvesting transport\cite{Collini2010, Plenio2008, Mohseni2008 }. In our case, it is the competition between the coherent and the dissipator parts what accelerates the convergence to the stationary distribution of walkers, {\it i.e.}, the ranking. Finally, in Fig. \ref{fig:tau}1.g we show the scaling of the ratio $\tau^{QR}/\tau^{PR}$ as a function of the size $N$ 
pointing out that the optimal enhancement is not a matter of the finiteness of the network, as the value $\tau^{QR}/\tau^{PR}\simeq 0.9$ remains roughly unaltered with $N$. In addition, Fig. \ref{fig:tau}1.g shows that classical and quantum walks belong to the same complexity class.  This latter result is quite expected as the hybrid walk can be continuously deformed into the fully classical one, by doing $\alpha \to 1$.  Since the convergence time relies on  the value of $\lambda_1$ (and it can be found as  a perturbation from the classical limit by making $\alpha = 1 - \epsilon$ with $\epsilon \ll 1$), by appealing continuity, there is not a change in the $N$-dependence for $\tau^{QR}$ as compared to $\tau^{PR}$.

\subsection{Solving PR degeneracies} 
\begin{figure}
\includegraphics[width=0.9\columnwidth]{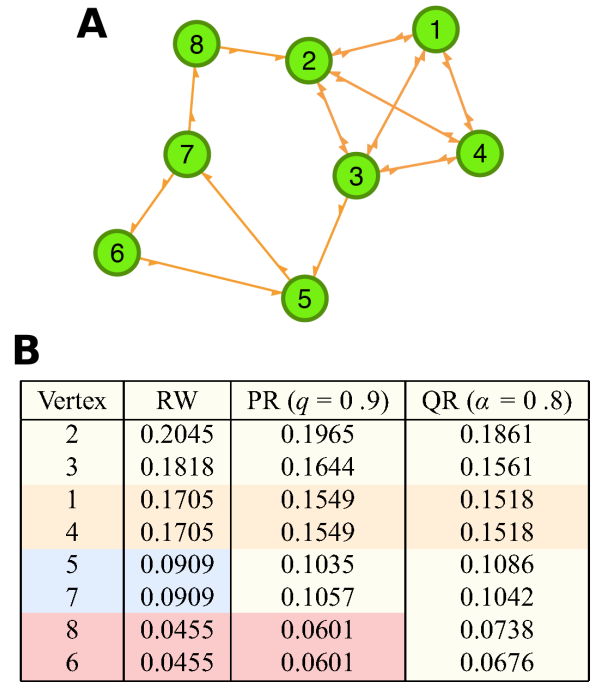}
\caption{{\bf QR {\it versus} classical ranks in a small graph.} In panel {\bf (a)} we show a simple directed graph composed of $8$ nodes. This graph can be briefly described as the sum of a densely connected (complete) subgraph (composed of nodes $1$, $2$, $3$ and $4$) plus a $3$-cycle (composed by the closed circuit $5\rightarrow 7\rightarrow 6\rightarrow 5$). These two structures are joined together by means of a link $3\rightarrow5$ and a directed path $7\rightarrow8\rightarrow2$. In panel {\bf (b)} we show the rankings obtained via two classical navigation schemes, the usual random walk (RW) and PR with $q=0.9$, and our QR. For the three navigation-based rankings the first $4$ nodes (composing the densely connected subset) are the most central ones, being headed by node $2$ (as it is the only one receiving extra flux from the peripheral part) and followed by node $3$ (being the only one with one outgoing link to the periphery). Note that nodes $1$ and $2$ play a similar role and thus they receive the same score regardless of the navigation method at work. The difference between the navigation schemes becomes more evident for the $4$ nodes in the periphery. RW assigns a similar rank to nodes $8$ and $6$ and the couple $5$ and $7$, thus the two last positions of the RW rank are degenerated. However, in these two cases, these degeneracies are not rooted in any similarity between the roles of the nodes. This problem is partially solved by PR which is able to unveil differences in the role of nodes $5$ and $7$ remaining, again, nodes $8$ and $6$ with the same score. Finally, QR solves the problem and assigns a different score to each of the $4$ peripheral nodes. Importantly, QR considers node $5$ as the most central element of the periphery, at variance with PR for which this position was occupied by node $7$. }
\label{fig:toy}
\end{figure}

After providing a reliable definition of QR that provides a unique classification of nodes, the fast convergence shown in the previous paragraph supports its feasibility as a navigation-based ranking method.  At this point the important issue relies on the novel features introduced in the ranking itself by the interplay of quantum and classical effects that QR incorporates. To perform a comparison between QR and PR we first rely on a simple network\cite{delvenne} in which a qualitative rank can be made at first sight to check the output of different rank algorithms. 
In particular, we will test the rank provided by three navigation dynamics: {\it (i)} the usual random walk, {\it (ii)} PR, and {\it (iii)} QR. 
The simple network is a directed graph composed of eight nodes (see Fig. 2.a) that can be described as the sum of a densely connected core (composed of nodes $1$, $2$, $3$ and $4$) plus a peripheral cyclic circuit (composed 
of nodes $5$, $6$ and $7$). Given the simple structure of this graph, it is easy to have some insight about the expected ranking. The nodes belonging to the densely connected core will accumulate most of the walkers' flow in the graph while, in the peripheral part, node $5$ will receive all the outgoing flow from the connected core. Therefore, these nodes should occupy the top positions of a proper ranking algorithm. 

In Fig. \ref{fig:toy}2.b we show a table containing the ranks produced by the aforementioned three navigation schemes. 
The three ranks show that the four nodes belonging to the densely connected core obtain the largest ranks, being 
node $2$ the most central node as it is the unique element receiving extra-flow from the peripheral part of the 
graph. On the other hand, nodes $1$ and $4$ obtain a similar centrality score highlighting their equal role within the graph. 
The differences between the three navigation schemes show up in the classification of the peripheral 
nodes. First, the classical random walk assigns similar scores to the couples of nodes $(5,7)$ and $(6,8)$. 
This degeneracy is a product of some local similarities between the two nodes of each couple. Nodes $8$ and $6$ allocate the same incoming flow of walkers (that coming from node $7$) while in the case of nodes $7$ and $5$ it is clear that all the flow reaching node $5$ from the connected core passes directly to node $7$. However, at a global scale, the role of these four peripheral nodes is not identical, thus challenging the reliability of the classical walk as a method to capture the differences between the global role of nodes.

The degenerate rank produced by the classical random walk demands the use of navigation algorithms with a broader information horizon so to unveil the differences in the role played by peripheral nodes at a global scale. PR incorporates this non-local ingredient by means of the long-range hops incorporated in the Google matrix (\ref{Gmatrix}). From Fig.\ref{fig:toy}2.b we observe the former degeneracies seems to be partially solved by PR since only nodes $6$ and $8$ remain with the same scores. 
%However, the solution to the degeneracy of nodes $5$ and $7$ is not satisfactory enough as PR considers node $7$ the most important of the peripheral elements. 
A more reliable ranking is provided by the non-local nature of QR. First, the two degeneracies are finally distorted so that each of the $4$ peripheral nodes have different scores. Interestingly, the quantum hierarchy among the peripheral nodes is headed by node $5$ followed by node $7$ (the most important node as ranked by PR)
whereas node $6$ (a simple connector node between nodes $7$ and $5$) remains as the less important node of the system.

\subsection{QR in real networks} 

\begin{figure*}
\includegraphics[width=1.95\columnwidth]{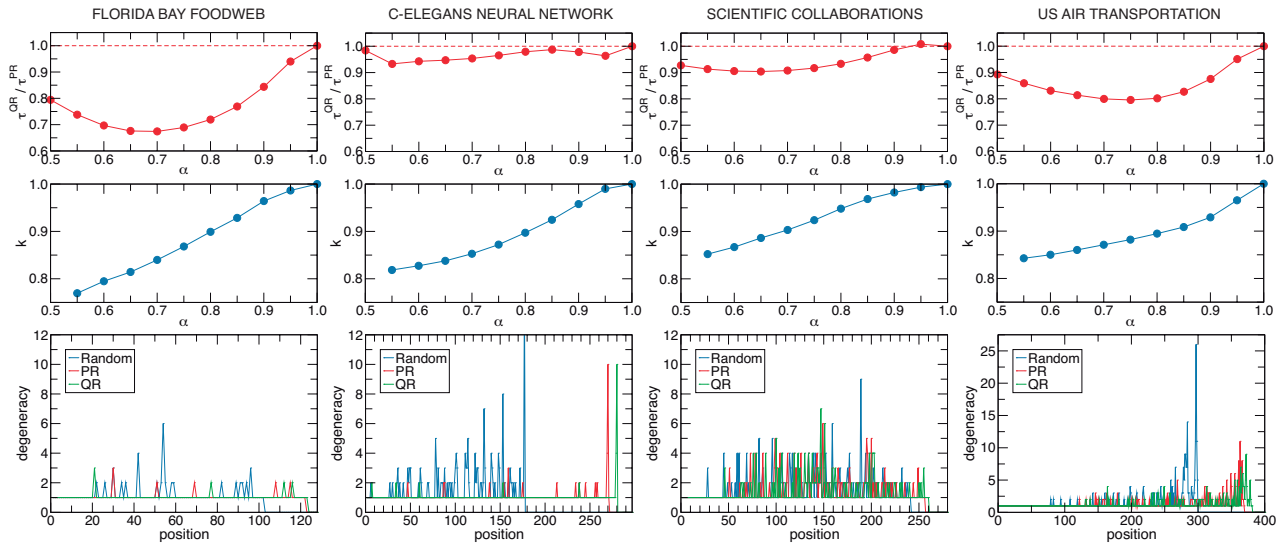}
\caption{{\bf Quantitative differences between the classifications of QR and PR.} In this figure we show the quantitative characterization of the performance of QR in real networks, namely (from left to right): the foodweb of the Florida bay ecosystem, the neural network of {\it C. Elegans}, a scientific collaboration network and the US air transportation system. In each of the top panels we show the evolution of the ratio between the convergence times of QR and PR, $\tau^{QR}/\tau^{PR}$, as a function of the quantum-classical interplay, $\alpha$. From these panels we observe, as in Fig. 1, that QR improves the convergence time of PR reaching  the maximum  outperformance when $\alpha^{opt}<1$. Those panels in the middle compare the rankings obtained by PR and QR by means of their Kendall coefficient, $K$, which takes a value $K=1$ for a perfect concordance between two classifications whereas $K=0$ for null agreement. As the panels show, $K$ decreases monotonously with $\alpha$ pointing out that the classifications change as quantumness is introduced in the QR. Finally, the bottom panels represent the degeneracy found in each of the positions of both QR and classical (RW and PR) classifications for the case $\alpha=0.9$. The fact that QR yields rankings with more positions implies that its ranks are less degenerated than those obtained via classical navigation schemes.}
\label{fig:real1}
\end{figure*}

To test the robustness of our previous findings it is convenient to implement the QR navigation in real-world networks in which more complex connectivity patterns appear. In particular, we have considered four real networks of different nature, namely, the foodweb of the Florida bay ecosystem\cite{baywet} ($N=128$), the neural network of {\it C. Elegans}\cite{watts,cen} ($N=297)$, the network of direct flights between the major commercial airports in the US\cite{usair} ($N=500$), and a network of scientific collaborations in the field of network science\cite{coll} ($N=379$).
%and their connections via direct flights\cite{usair}. 
%In these cases the key point is to detect those elements that play a central role in the complex organization of these real networks.

First we revisit the effects that the quantum-classical interplay of QR causes on the time of convergence to a stationary ranking. 
In the top panels of Fig. 3 we have plotted the ratio $\tau^{QR}/\tau^{PR}$ as a function of the classical-quantum interplay $\alpha$. The curves confirm the improvement in the convergence times shown in Fig. 1 when a moderate quantum interplay is present in the QR navigation. In addition to this enhancement, we observed in Fig. 2 that the introduction of quantumness produces relevant changes between the classifications obtained with PR and QR. For large networks, the comparison between QR and PR is not as straightforward as for the simple graph of Fig. 2. However, a coarse-grained evaluation of the difference between QR and PR can be obtained by means of the Kendall coefficient\cite{kendall}, $K$, which takes a value $K=1$ for perfect concordance between two rankings while $K=0$ for null agreement. The panels in the middle if Fig. 3 show the value of the Kendall coefficient between the classifications of QR and PR as a function of the quantum-classical interplay of QR, $K(\alpha)$. We observe that, in all the cases, the values of $K$ remain large for the values of $\alpha$ in which QR outperforms PR, but small differences appear as soon as quantum effects enter into play.

Having quantified the differences that QR introduces in the final ranking, we now explore in more detail the nature of the changes. First we tackle the problem of degeneracies to test if, as shown in Fig. 2, QR provides a classification without less degenerate positions than classical rankings. In the bottom panels of Fig. 3 we show the number of nodes occupying each of the positions contained in classical rankings (RW and PR) and QR. These comparisons reveal that, again, QR splits some of the degeneracies found in the classifications obtained by classical means. 
%This result is more obvious from the inset of the figure as the classification of QR spans around $10$ positions more than that obtained by PR. 
Thus, also in large networks, quantumness produces a finer discrimination of nodes centralities.
 
Apart from solving degeneracies, the differences between QR and PR unveiled by the Kendall coefficient of both rankings raise the interesting question of which nodes are affected when introducing quantum effects into the PR navigation. In Figure 4.a we show, for the case of the US air transportation network, the difference between the position of a node in the QR and the one assigned by PR as a function of the position in the first classification. The nodes ranked in the top part are not affected while most of the rises and falls are concentrated in the medium and lowest part of the ranking. This is, again, in agreement with the results found in the small graph in Figure 2. However, the case of the air transportation network is far more complex than that of the simple toy graph and we can gain more insight about the positive and negative effects that QR causes in the classification of lowly connected nodes.

In Figures 4.b-d we plot the US air transportation network by coloring each node according to the sign of  the difference between its QR and PR rankings. Those nodes in red have fallen in the QR (with respect to their position in the PR) whereas, for those nodes in green, QR has improved their PR score. Finally, those nodes being roughly at the same position for both PR and QR are colored in yellow. First we notice that the structure of the network is governed by a set (central part of the plot) of highly connected nodes (the hubs) whose degree is remarkably large. This is the set of large-score nodes whose elements do not suffer any relevant change in their position. Additionally, we have a set of lowly connected nodes whose connections are mainly directed to the largest hubs. Most of these nodes are colored in red since they have been relegated to lower rank positions by QR. The fall of these nodes is balanced by the raise in QR of another set of lowly connected nodes (colored in green) located in the periphery of the network. In Figure 4.e. we analyze this effect in more detail. The nodes having the largest improvements in their ranks are those connected to neighbors with large QR and moderate degree, while the nodes that drop in the ranking are those that are connected to hubs distributing their influence amongst a large number of connections. Thus, QR ranks low degree nodes as a product of their global importance rather than to their local proximity to the most important elements of the system.

\begin{figure*}
\includegraphics[width=1.95\columnwidth,angle=0]{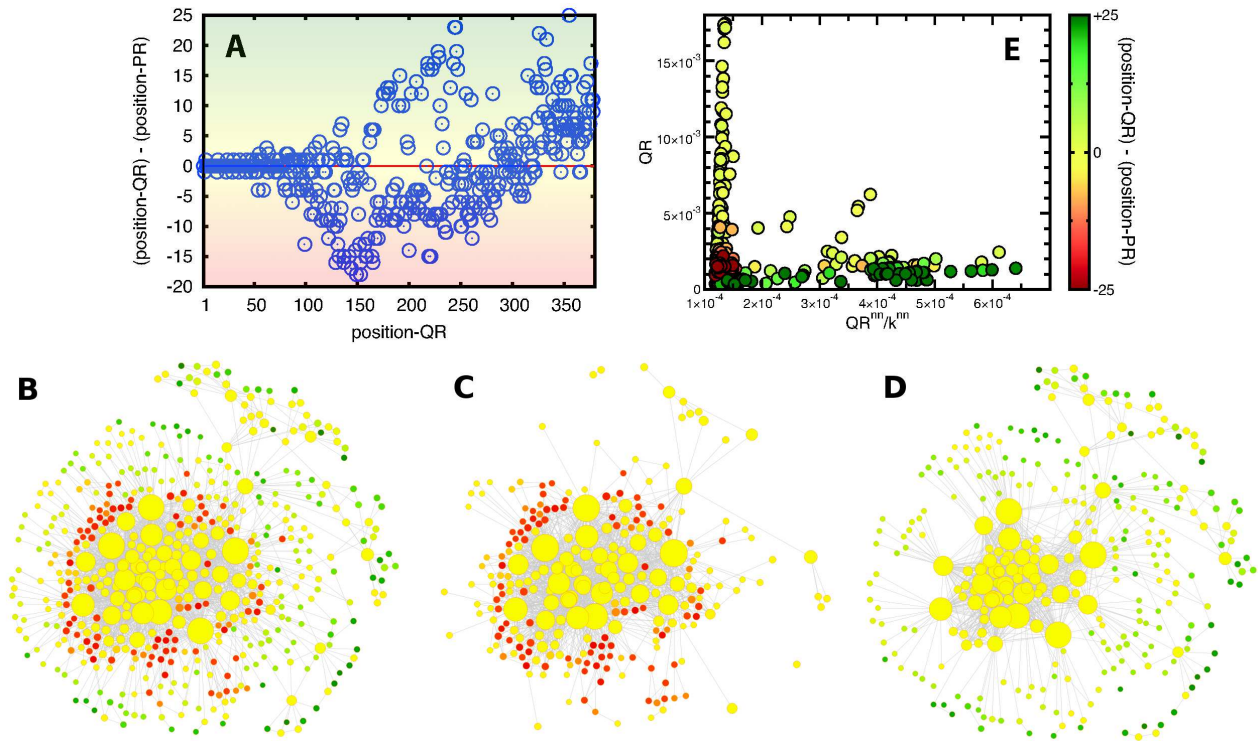}
\caption{{\bf Qualitative differences between the classifications of QR and PR.} Panel {\bf (a)} shows, for each node in the US air transportation network, the difference between the position assigned by QR and that obtained via PR as a function of the first one. It is clear that for the most central nodes there are few differences between the position assigned by both navigation schemes. However, for nodes with moderate and low centrality the differences between QR and PR show up. In panel  {\bf (b)} we plot the US transportation network. The size of the nodes is proportional to their QR score whereas they are colored in green, yellow and red when the difference between QR and PR positions is positive, zero or negative respectively. In panel {\bf (c)} and {\bf (d)} we plot again the network but filtering those nodes with positive and negative values for the former difference respectively. To gain more insight, in panel {\bf (e)} we plot the QR of each node as a function of the average QR of its nearest neighbors, $QR^{nn}$, divided by their average degree, $k^{nn}$. The color of each point denotes the difference in the positions assigned by QR and PR.
From this panel we observe that those points in green spans along the horizontal axis denoting that the nodes that are favored by QR are those connected to nodes with large QR but moderate connectivity [as it can be visualized from panels {\bf (b)}, {\bf (c)} and {\bf (d)}]. However, those points in red are localized in the bottom-left corner of the figure pointing out that those nodes punished by QR are connected to extremely large degree nodes (hubs) as, again, can be easily visualized in panels {\bf (b)}, {\bf (c)} and {\bf (d)}. Finally, those nodes that are unaffected by QR (yellow points) correspond to large QR scores.} 
\label{fig:real2}
\end{figure*}

\section*{Discussion}

In summary, we have proposed a new way to navigate complex networks via quantum walkers. Our idea extends classical navigation processes to the quantum domain, incorporating  coherences in the navigation through an otherwise classical set of nodes. We have first introduced a framework for the walkers motion that combines the quantum unitary evolution with the irreversible character of the classical walk. This hybrid navigation provides a unique stationary distribution for the occupation probabilities that allow to compute a reliable quantum rank. 

We have shown that the interplay between quantum and classical features in the random walk dynamics
produces a decrease in the number of navigation steps needed
%significant 
to converge to the stationary state of the walkers motion, thus making feasible the computation of the final rank by means of a physical navigation of a network. We emphasize that QR can be readily implemented without the need of any quantum hardware, which does not exist yet, 
%the computation of the QR is made by using a {\it classical computer}, {\it e.g.} propagating the quantum walk with the standard power method. Thus, QR can be readily implemented without the need of a quantum hardware, which does not exists yet, 
going beyond the current paradigm of quantum algorithms.

The introduction of quantum rules in the walkers also manifests the effect of quantum coherences in the properties of the rank itself. First we have shown that QR resolves undesired degeneracies found in ranks obtained with classical meanings. These degeneracies mainly affect those lowly connected nodes and it is within this subset where QR reveals more differences with its classical counterparts. Whereas those elements occupying the top part of the classical rank remain as the most central ones for QR, the rest of elements are subjected to important variations. In particular, we have observed that those elements which are connected to moderately connected nodes increase their popularity with QR while those nodes connected to the hubs of the network decrease their importance. Thus, quantumness seems to punish the fact of being connected to influential nodes when this link becomes diluted by the large popularity of the hub. This punishment is balanced with the increase of the status of those nodes being connected to moderately central ones. Concisely, this result unveils the quantum solution to a recommendation dilemma: quantumness sets nodes {\it recommended} by locally important elements apart from those {\it recommended} by hubs which, in its turn, {\it endorse} a huge amount of leaves.

As mentioned above, our algorithm does not rely on a {\it quantum implementation}. Alternatively to the efforts for solving {\it classical} complex problems by means quantum computers, {\it e.g.} finding the protein native state by means of quantum annealing\cite{Perdomo2012}, our interest here relies on the new features that quantum rules bring out when solving classical problems, {\it e.g.} exploring and ranking complex networks. On the other hand, the possible implementation of a quantum navigation method, such as ours, in a quantum machine will open the path to a more efficient application of quantum rank techniques.  To this aim, and considering the recent advances in quantum annealing protocols, one should address the optimal mapping of the network nodes into the (minimal) number of qubits, considering the complex network topology of the interactions between them.  

In addition,  our work is also 
complementary to recent studies on quantum complex networks focusing on the development of a {\it quantum internet}\cite{Kimble2008}. The backbone of these quantum networks is composed of photon emitters and receivers with the aim of developing a new  technology paradigm based on secure quantum cryptography protocols\cite{Galindo2002}. Obviously, as the number of nodes and connections increases, the quantum information community meets complex networked structures. This synergy has lead to the emergence of an exciting new field in which classical complex phenomena such as percolation\cite{Acin2007, Cuquet2009} or the small-world effect\cite{Perseguers2010, Wei2011} are revisited and addressed together with the new quantum machinery underlying these networks. Here we have opened the door between quantum and complexity sciences from the other side. We have considered the actual network configurations and applied a quantum perspective to the solution of a classical complex network problem, trying to understand whether quantum mechanics may improve the solution to open problems of inherent complex nature\cite{Fleming2011}.

%%%%%%%%%%%%%%%%%%%%%%%%%%%%%%%%%%%%%%%%%%%%%%%%%%%%
%%%%%%%%%%%%%%%%%%%%%%%% methods %%%%%%%%%%%%%%%%%%%%%%%
%%%%%%%%%%%%%%%%%%%%%%%%%%%%%%%%%%%%%%%%%%%%%%%%%%%%

\appendix

\section{Methods}

\subsection{Markovian Quantum Master equation and Random Walks in Complex networks}

Here, we discuss the theoretical framework for our algorithm.  We introduce the Markovian quantum master  equation (MQME) that generalizes its classical counterpart into the quantum domain.  We demonstrate that for Google type transition matrices, the MQME is relaxing, {\it i.e.}, it has always a unique stationary state.  Finally we tackle the convergence times for reaching such stationary solution.

To start with, we rewrite the master equation (\ref{crw}), for a classical Markovian random process:
\begin{equation}
\label{crw-m}
\frac{d}{dt} p_i = \sum_j M_{ij} p_j
\end{equation}
with $M = G - \mathbb {I}$, being $G$ the Google matrix.  The usual way of quantizing such random walk consists in introducing the Shr\"odinger equation $d_t |\psi \rangle = -{\rm i }/\hbar H |\psi \rangle$, where the occupation probabilities $p_i$ are replaced by the components $\langle i | \psi \rangle$ of the wave function, such that,
\begin{equation}
\label{qw}
\frac{d}{dt}
\langle i | \psi \rangle = - \frac{\rm i}{\hbar} \sum_j\langle i | H | j \rangle \langle j | \psi \rangle
\; .
\end{equation}
To finish the identification with (\ref{crw-m}) one sets $\langle i | H | j \rangle = M_{ij} $.  At this point, it is mandatory to recall that $H$ is an Hermitian operator, namely $H = H^ \dagger$, implying that $ \langle i | H | j \rangle  = \langle j | H | i \rangle$ (the matrix $M$ is real). Therefore the above procedure is restricted to  undirected graphs, $M_{ij} = M_{ji}$.   Similar problems arise with the discrete (coined) version of random walks\cite{Kendon2006}.  To solve the problem of directionality one can resort to the Szegedy recipe\cite{Szegedy2004}.  However, in any of the above cases, the evolution is unitary. 
In other words, classical-quantum identification is not a straightforward task since classical Eq. (\ref{crw-m}) is irreversible while those quantum ones are reversible\cite{Paparo2011}.

There is a way of introducing both irreversibility and quantum coherences in a close way as it was recently pointed\cite{Aspuru}. The idea is to work with Markovian master equations for the density matrix $\varrho$ rather than with the aforementioned Schr\"odinger equation.  Being definite, the general theoretical framework is based in a time local master equation,
\begin{equation}
\label{dr}
\frac{d \varrho}{dt} =
{\mathcal L} \varrho
\end{equation}
with $\mathcal L$  being a differential operator encapsulating the {\it standard form} in which any Markovian master equation can be casted in\cite{Rivas},
\begin{align}
\label{lindblad-m}
\mathcal L  \varrho= &-{\rm i}(1-\alpha)[H,\varrho]
\\ \nonumber
&+\alpha\sum_{(i,j)} \gamma_{(i,j)}\left(L_{(i,j)}  \varrho L_{(i,j)}^\dagger-\frac{1}{2}
\{L_{(i,j)}^\dagger L_{(i,j)},\varrho \}\right)\;,
\end{align}
where $\gamma_{(i,j)} >0$ and  $L_{(i,j)}$ are linear operators, acting on a finite dimensional linear (Banach) space: $\mathcal {B(H)}$. The two-index notation ${(i,j)}$ will be clear now.
Equations (\ref{dr}) and (\ref{lindblad-m}) yield an irreversible dynamics where the directionality of links can be readily implemented.   In order to match  classical and quantum random walks, we write the set of equations for the diagonal elements of the density matrix, $\varrho$. Let us consider $\alpha = 1$ in (\ref{lindblad-m}) and  $L_{(i,j)} = |i \rangle \langle j |$. Then, 
\begin{equation}
\label{rnnd}
\frac{d}{dt} \varrho_{ii} = \sum_j  (\gamma_{ij} - \delta_{ij}) \rho_{jj} \; .
\end{equation}
In this case the equations for the non-diagonal elements,  $\varrho_{ij}$ ($i \neq j$) become decoupled from the diagonal ones.  By simple inspection  of (\ref{crw-m})  and recalling $\varrho_{ii} = p_i$  we identify $\gamma_{ij} = G_{ij}$ so that both the classical equations (\ref{crw}) and the latter (\ref{rnnd}) match.
It is worth mentioning that, for $\alpha =1$, the non-diagonal elements decay exponentially in time, thus they do not contribute to the stationary state.  Denoting $\varrho^*$ the stationary state, such that $d_t \varrho^* = 0$, it is easy to show (see below for a deeper discussion) that any initial state will asymptotically approach to $\varrho^*$.  Therefore in the discussed limit so far ($\alpha=1$)   the stationary state has the  non-diagonal elements zero, $\varrho^*_{ij} = 0 \; i \neq j$, while the diagonal elements are equal to the stationary distribution of the classical random walk, $\varrho_{ii}^* = p_i^*$.

The latter case is a trivial extension of the classical navigation process, without any quantum ingredient for the ranking. In general, we will  deal with $\alpha \neq 1$.  We choose for the Hamiltonian,  $H_{ij} = 1$ if the nodes $i$ and $j$ are connected and zero otherwise,   in complete analogy to the identification used in quantum walks. It should be noted here that the hamiltonian {\it mixes} both diagonal and non-diagonal elements making the dynamics non-trivial and different from the  classical one.  Indeed, going to the limiting case of $\alpha = 0$ one finds
\begin{equation}
\label{rnnc}
\frac{d}{dt} \varrho_{ii} = -{\rm i }  \;  \sum_j 2 H_{ij} ( \varrho_{ji} - \varrho_{ij})
\end{equation}
where the diagonal and non-diagonal elements are coupled. 
%When convining equations (\ref{rnnd}) and (\ref{rnnc}) with their respective weights $\alpha$ and $1- \alpha$ respectively a rich dynamics emerges  together with a family of novel stationary states depending on the $\alpha$ value. 

\subsection{Existence and uniqueness of the stationary solution}

We  now prove a key result for our purposes: the existence and uniqueness of the stationary solution in the quantum random walk, $\varrho^*$. In order to do so, we start with the following definition:

{\it 
A master equation  (\ref{dr}) with $\mathcal L$ given in (\ref{lindblad-m}) is relaxing if there exists a unique (steady) state $\varrho^*$ such that any initial state $\varrho_0$ converges to $\varrho^*$ for sufficiently large times.
}

Besides, we recall a Theorem shown by Spohn\cite{Spohn}:

{\it
The master equation (\ref{dr}) with $\mathcal L$ given in (\ref{lindblad-m}) is relaxing if (i) the set $\{L_{(i,j)}\}$ is self-adjoint (the adjoint of every $L_{(i,j)}$, namely $L_{(i,j)}^\dagger$, is inside the set) and (ii) the only operators commuting with all the $L_{(i,j)}$ are proportional to the identity.
}
 
By putting all together, we state:

{\bf Theorem}.---
{\it A quantum random walk with master equation (\ref{dr}) and $\mathcal L$ given in (\ref{lindblad-m}), where $L_{(i,j)} = |i\rangle \langle j |$ and $\gamma_{ij} = G_{ij}$ being $G$ the  Google matrix  (\ref{Gmatrix}) is relaxing, {\it i.e.}, it  has always  a stationary state and it is  unique.
}

{\bf Proof}.---
We first show that the set is self-adjoint.  Given $L_{(i,j)} = |i \rangle \langle j|$, its self adjoint is $L_{(i,j)}^\dagger = |j \rangle \langle i | = L_{(j,i)}$.  We notice that the Google matrix links the nodes all-to-all. Therefore $G_{ij} \neq 0$ $\forall \, i,j$ providing that the adjoint operator is also in the set. Finally, be  $A$ an operator on $\mathcal {B(H)}$, with decomposition: $A = \sum a_{kl} |k\rangle \langle l |$. Make the commutator with $L_{(i,j)}$:
\begin{widetext}
\begin{equation}
\nonumber
[A, L_{ij}] = \sum_{k, l} a_{kl} \big [ \;  |k\rangle \langle l|, |i \rangle \langle j| \; \big ]
=
\sum_{l \neq i, j}  (a_{li} |l\rangle \langle j| - a_{jl} |i\rangle \langle l|) +(a_{ii} - a_{jj}) |i\rangle \langle j| + a_{ij}|j\rangle \langle j | - a_{ij}^* |i \rangle \langle i |\;,
\end{equation}
\end{widetext}
since this commutator  should be zero for every $L_{(i, j)}$ the only possible solution is that the non-diagonal elements of $a_{ij}$ ($i\neq j$)  vanish and its diagonal ones are equal, $a_{ii} = a_{jj}$.  Therefore, $A$ is proportional to the identity and, consequently, we met the two conditions in the Spohn theorem, proving our result.

\subsection{Approach to the stationary solution}
Being  $\mathcal L$ time independent, the formal solution of (\ref{lindblad-m}) is
\begin{equation}
\varrho (t) = {\rm e}^{{\mathcal L} t} \varrho (t_0)
\; ,
\end{equation}
given any initial state $\varrho (t_0)$. It turns out that ${\mathcal L}$ can always be decomposed into a direct sum of Jordan forms\cite{Rivas}: $S^{-1}{\mathcal L} S = \mathcal L^{(0)} \oplus \mathcal L^{(1)} \oplus ... \oplus \mathcal L^{(K)} $, where,
\begin{equation}
 \mathcal L^{(k)}
 =
 \left (
 \begin{array}{ccccc}
 \lambda_k & 1 & 0 &  \cdots &0
 \\
 0 & \lambda_k & 1 & \ddots & 0
 \\ 
 \vdots & \ddots & \ddots & \ddots & 0
 \\
 \vdots &  & \ddots & \ddots & 1
 \\
 0 & \cdots & \cdots & 0 & \lambda_k
\end{array}
 \right )
\end{equation} 
and $\mathcal {L}^{(0)} = (1)$, {\it i.e.} a $1 \times 1$ matrix with entrance 1, corresponding to the stationary solution ($\mathcal L$ relaxing) . 
Written in this way, the exponential yields 
${\rm e}^{\mathcal {L}t} = S (1) \oplus ({\rm e}^{\lambda_1 t} {\rm e}^{{\bf N}_1 t})\oplus  ... \oplus ({\rm e}^{\lambda_K t} {\rm e}^{{\bf N}_K t}) S^{-1}$, with ${\bf N}_k$ nilpotent matrices, so that the convergence to the unique stationary solution is guaranteed since,
$ \rm {lim}_{t \to \infty} ={\rm  e}^{\mathcal L t} = S ( 1 \oplus 0 \oplus ... \oplus 0 )S^{-1}$, see\cite{Rivas}.  The convergence ratio depends on how fast the blocks ${\rm e}^{\lambda_k t} {\rm e}^{{\bf N}_k t}$ approach to zero, so that it is bounded by the value of the largest eigenvalue (different from zero) $\lambda_1$. This Jordan-Block evolution resembles the classical case, where the power method is widely used. Similar procedure could be used here by splitting the evolution in discrete time steps: ${\rm e} ^{{\mathcal L} t} = \Pi_{m=1}^M {\rm e} ^{{\mathcal L} \delta}$ with $M \delta = t$. Alternatively, one could study the convergence by means of other methods, {\it e.g.} the  proposed  quantum adiabatic algorithms\cite{Garnerone2011} . 

We choose to integrate numerically the differential form until the the stationary solution is reached.  We define the convergence time $\tau^{QR}$ such that for $t > \tau^{QR}$ then  $|| \varrho (t) - \varrho^* || < \epsilon$, with the norm $ || \varrho || = \sqrt {\sum_i (\varrho_{ii}(t) - \varrho_{ii}^* )^2 }$.  To compare both PR and QR we take the ratio
 $\tau^{QR} / \tau^{PR}$, where $\tau^{PR}$ is the convergence time, integrating the classical random walk, or equivalently the quantum master equation when $\alpha = 1$ see (\ref{rnnd}).   As initial condition we choose $\rho_{ij} =  \delta_{ij}/N$.

As a final comment, we highlight that working with density matrices, $\varrho$, involves, in principle, the integration of $N^2 \times N^2$ elements. Thus, the computational cost per integration step becomes large, as compared with the classical case where matrices involved are $N \times N$.  However, it is possible to overcome this issue by working within the framework of the quantum stochastic Schr\"odinger equation that, being equivalent to our master equation, allows to recover matrices of size  $N \times N$ as in the classical case\cite{VanKampen}.

%\section*{References}

%\begin{addendum}

% \item[Competing Interests] The authors declare that they have no competing financial interests.

 %\item[Correspondence] Correspondence and requests for materials should be addressed
% to J.G.G.~(email: gardenes@gmail.com).
%\end{addendum}

\section*{Acknowledgments}

This work has been partially supported by the Spanish DGICYT under projects FIS2009-13730-C02-02, FIS2009-13364-C02-01, CSD2007-046-Nanolight.es, MTM2009-13848, FIS2011-25167 and EXPLORA FIS2011-14539-E; by the Generalitat de Catalunya and the Comunidad de Arag\'on through Project No. 2009SGR0838, FMI22/10; and by the J.S. McDonnell Foundation Research 344 Award. J.G.G is supported by MICINN through the Ram\'on y Cajal program

%\section*{Author Contributions}
%D.Z., J.D. and J.G.G devised the model and designed the study. E.S.B and J.D. carried out the numerical simulations. D.Z., J.D. and
%J.G.G. analyzed the data and prepared the figures. D.Z., E.S.B., J.D. and J.G.G. wrote the main text of the manuscript.

%\section*{Correspondence} Correspondence and requests for materials should be addressed
 %  to J.G.G.~(email: gardenes@gmail.com).

%\section*{Additional Information}
%\subsubsection*{Competing Financial Interests} The authors declare no competing financial interests.

\end{document}